# Investigation of alumina formed by plasma electrolytic saturation

**Seyad Mohammad Mousavi Khoie, Mohammad Falaki, Babak Ghorbanian***

*Corresponding Author Address: Department of Metallurgy and Materials Science, Semnan University, Iran.
ghorbanian.babak@semnan.ac.ir*

**Abstract**

The process of alumina coating on the steel surface plasma electrolytic method is a method for increasing the corrosion resistance, in this process, electrolytic aluminum containing compounds should be able to cover the surface. The results show the size of a micron coating of cauliflower shape and distance of an average of about 6 microns. The coating on steel sample acted as a passive layer and corrosion resistance increases slightly. This is due to the low corrosion resistance because of the lack of uniformity and surface finish are not.

**Keywords**: Plasma electrolytic saturation coverage of alumina, steel St12

**Introduction**

The phenomenon of surface engineering in electrolytic plasma can be named under the general title of plasma electrolytic deposition (PED), which includes two branches with the titles of oxidation by the electrolytic plasma method (Plasma Electrolytic Oxidation, PEO) and saturated electrolytic plasma. (Plasma Electrolytic Saturation, PES), although the discharge in the electrolysis process, more than a century ago by S. Luginovo was discovered and studied in detail in the 1930s by Günther Scholtz and his colleagues, but its practical aspects were discovered in the 1960s when Makham.C. Neal and Gross used electric discharge to deposit on the surface of a cadmium piece under arc discharge conditions, it was developed. And in the mid-70s, the advanced anodizing process by two Russian scientists named J. It was developed by A. Markov and JV Markov. In the 1980s in Russia, micro-arc or electric discharge in the process of oxide deposition to be applied on the surface of various metals was investigated by Yerokhin and his colleagues. During the 1980s, new situations were created for the industrial application of the technique known as saturation in the electrolyte plasma environment (PES). However, the further development of the processes requires a better understanding of the chemical and physical principles of the phenomenon of plasma on the electrode during electrolysis [1-4].
Penetration of elements by plasma can be obtained as a result of electric discharge with direct current or electric discharge with short waves. It should be noted that in the penetration of elements with plasma, the target or sample is generally placed inside the plasma or on the surface of the sample boundary. This issue is especially important in the construction of thin layers.
Due to the application of electrical conductivity, the rapid increase in voltage causes the creation of many bubbles on the surface of the working sample, due to which, the electrolyzed solution of boron ions in the electrolyte, which are in contact with the working sample, turn into steam. Up to about 75 V, the voltage-current characteristic of the cell is in accordance with Ohm's law. When the voltage exceeds this range, a spark is created on the surface, these sparks cause the boron bubbles to burst, and due to their high temperature, the gas in these bubbles becomes free radicals. This environment containing plasma radicals is called. An increase in the applied voltage causes the sparks to intensify. When the voltage reaches 175 volts, the sparks around the electrode become a continuous ignition. The complete separation of the electrode from the electrolyte by a continuous plasma coating causes a sharp decrease in the current of the electrode surface and the surface of the working sample is bombarded by high-energy radicals. It penetrates to the surface of the part or is oxidized by the oxygen inside the bubble and covers the surface [4]. In this research, it has been tried to create an alumina coating on the surface by oxidizing the aluminum element. Figure 1 shows a schematic view of the electrolytic plasma process.

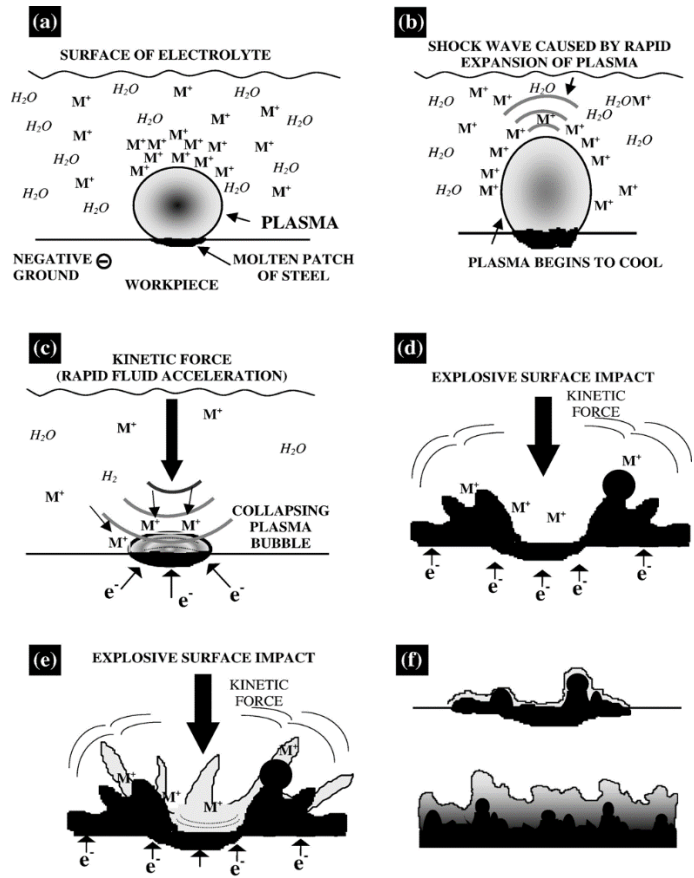

Figure 1. Schematic of the PES process[7,3]

**Research materials and methods**
The samples tested in this research are plain carbon steel St12, the results of quantum analysis are shown in Table 1.

Table 1. Analysis of the chemical composition of steel

| Sn | Co | Cu | Mo | Ni | S | P | Mn | Si | C |
|---|---|---|---|---|---|---|---|---|---|
| 0.006 | 0.004 | 0.035 | 0.005 | 0.030 | 0.007 | 0.007 | 0.200 | 0.010 | 0.050 |

Sheet steel samples with dimensions of 2 x 1 cm were cut into the required number with a cutter. Then the samples were polished with 80 and 400 grit sandpaper.

The components of the electrolyte solution must contain the aluminum element and be ionized into positive aluminum ions due to the plasma process. The selected electrolyte is obtained by dissolving aluminum metal (7 grams) in 37% hydrochloric acid (50 ml) along with sodium chloride (50 grams with 99% purity to increase the conductivity) in two liters of water. The conductivity was measured with a conductivity meter, and its numerical value was 47 millisiemens (ms).

In order to see the surface topography and map from the SEM scanning electron microscope, and to check the corrosion of the sample, a polarization test was performed on the sample.

Polarization test was used to investigate the behavior of corrosion resistance. This test was performed by the Autolab-pgstat302n device with a scan rate of 1 mV/s with a potential range of ±100 mV compared to the open circuit potential (OCP) in a solution of 3.5% by weight of sodium chloride (Merck company) for 15 minutes. The maximum current in the software was set to 1 amp and the minimum to 1 micro amp. This test was performed in a three-electrode cell, which were the working electrode, the reference electrode, and the auxiliary electrode, respectively: the coated sample (uncoated steel sample), the Ag/AgCl electrode, and the platinum electrode. For these tests, an area of 1 square centimeter of the steel surface was covered. was exposed to the electrolyte and the rest was covered with a molten mixture of bisox. This test was done at ambient temperature (25±5 degrees Celsius). The immersion time of the working electrode was considered to be 24 hours before starting the experiment in order to reach a stable state. Corrosion current density (icorr) was obtained by Tofel extrapolation technique at ±100 mV around the open circuit potential. Nova 1.6 software was used to evaluate the polarization test results.

**Results and discussion**

Figure 2 shows the X-ray diffraction diagram. As shown in the figure, in addition to the iron peak, aluminum oxide peaks can also be seen on the surface of the sample, which indicates that there is a coating of alumina on the surface of St12 steel. In other words, it can be said that the coating created on the surface of St12 is made of alpha aluminum oxide and it is expected to increase the corrosion resistance. One of the most important characteristics of the coatings obtained from electrolytic plasma is the lack of uniformity and roughness of the coating, which causes unwanted separations and then creates noises in X-ray diffraction.

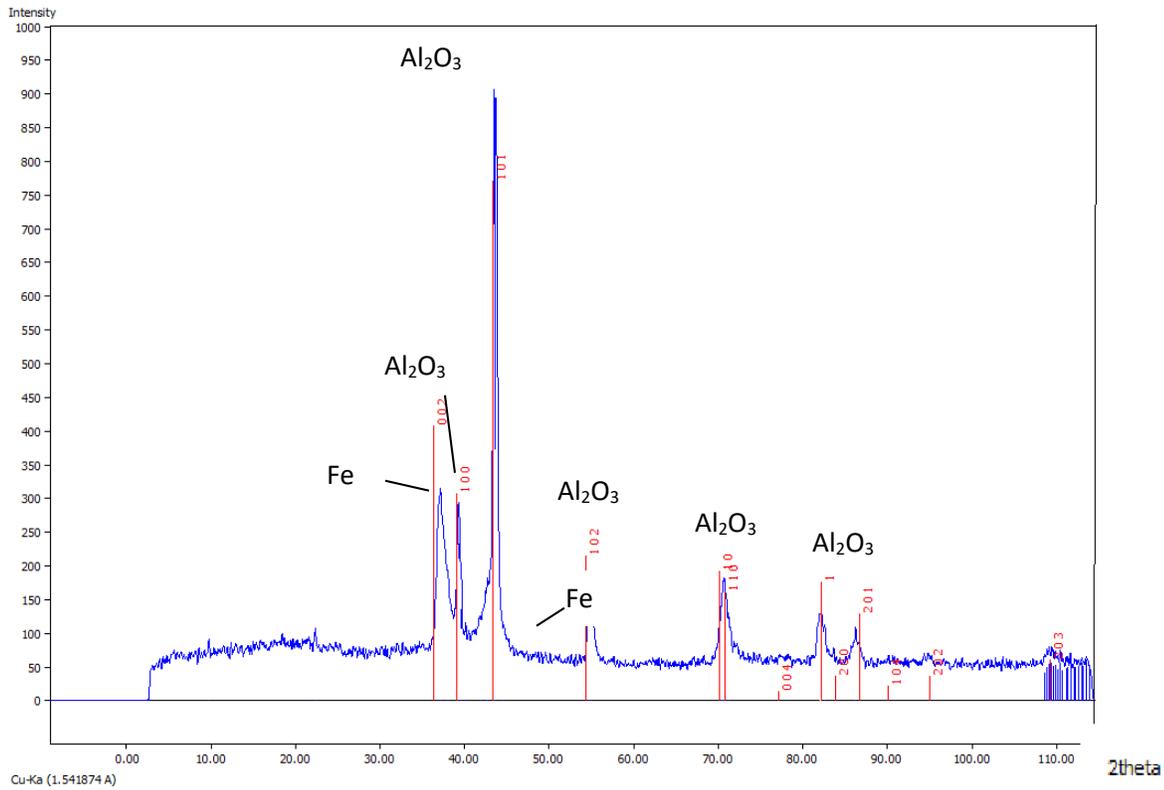

Figure 2. X-ray diffraction obtained from the tested sample

Figure 3 shows the scanning electron microscope image of the surface of the alumina coating created on the steel surface. As shown in Figure 3(a), the resulting coating has a cauliflower-shaped morphology. According to the theory of plasma coatings, these coatings are created by applying an electric field and sparking. Therefore, it is predicted that the cover created by seed germination due to a spark and then creating another bud on top of the previous one will have cauliflower morphology. Figure 3 (a) is a scanning electron microscope image at a magnification of 1000 times, as the figure shows, the distance between cauliflower seeds is about 6 microns. According to Figure 2 (b), the size of a cauliflower particle is about 1-2 microns.

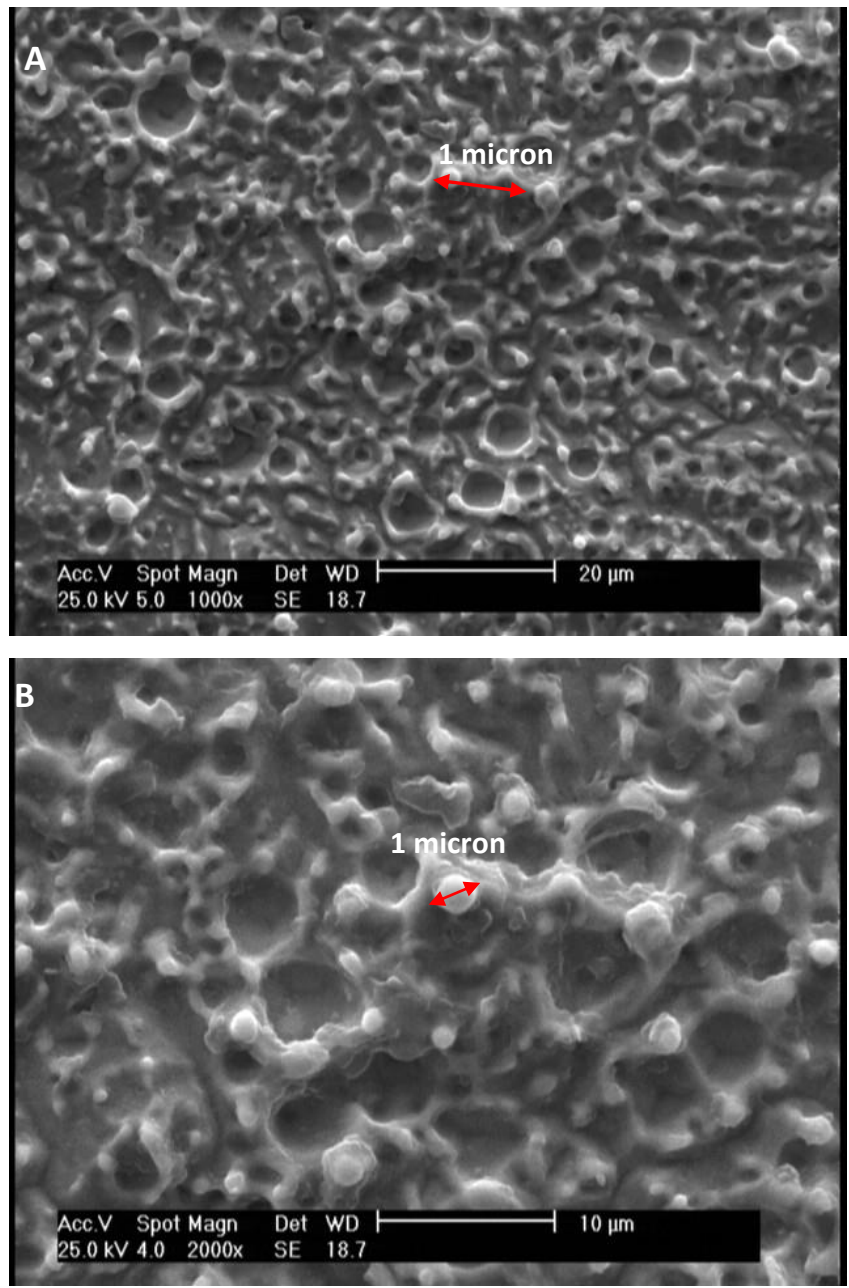

Figure 3. Scanning electron microscopy a) 500x magnification b) 1000x magnification

Figure 4 shows the image of the scanning electron microscope from the edge of the sample, as shown in Figure 4(a), the size of cauliflower particles at the edge of the sample has a larger grain size, and as we move away from the edge of the sample, the size of the grains decreases. Figure 4(b) is the scanning electron microscope image of the edge of the sample, while Figure 4(c) shows the electron microscope image of the center of the sample. As it is clear from Figures 4 (b) and (c) that the size of cauliflower particles at the edge of the sample is about 7 microns, while the size of the particles in the center of the sample is less than 1 micron. The reason for this increase in cauliflower seeds at the edge of the sample is due to the lower cross-sectional area and the increase in the current at the edges, which increases the volume of sparks at the edges of the sample. This increase in the volume of sparks increases the amount of ionization of the aluminum element and during that the alumina grains become larger.

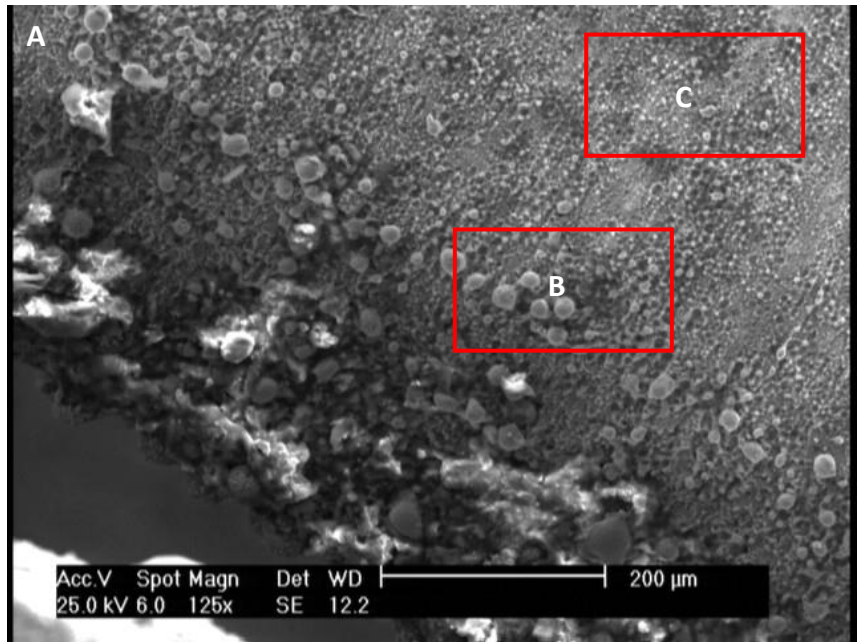

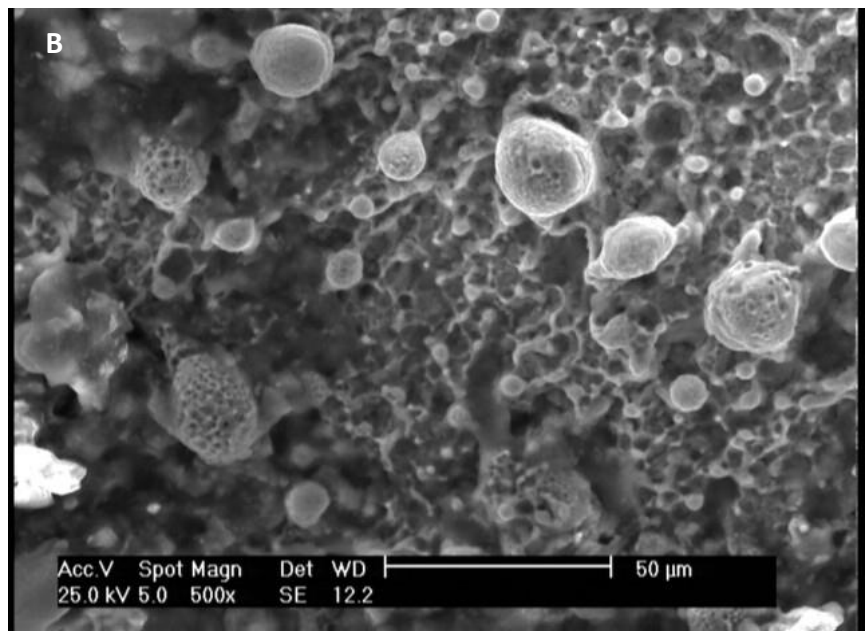

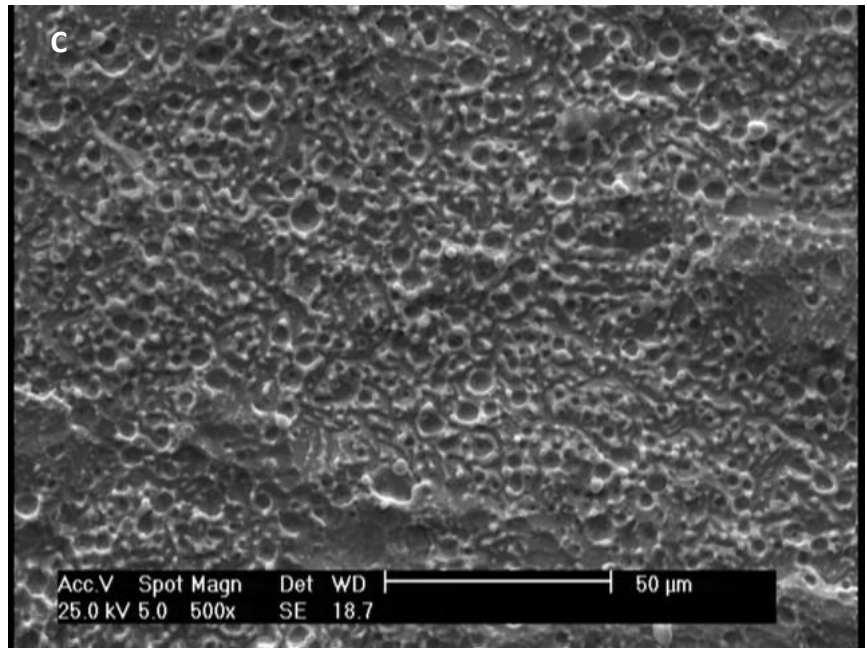

Figure 4. Scanning electron microscope image a) edge and center of the sample b) edge of the sample c) center of the sample

In this experiment, the determination of the corrosion rate and mechanism of low carbon steel in contact with the corrosive environment of sea water is considered (the schematic view of which is shown in Figure 5).

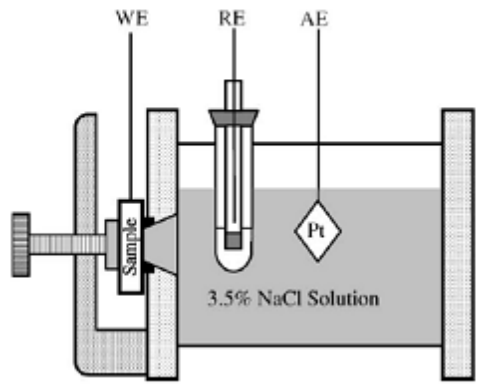

Figure 5. Schematic of the standard cell of the potentiostat device and how the sample is placed in it.

As Figure 6 shows, the sample with an oxide coating has a higher corrosion current, which indicates an increase in the corrosion rate. According to the Purbeh curve for aluminum, the alumina coating acts as a passive coating in the corrosive conditions with the pH of salt solutions and increases the corrosion resistance. In this case, the aluminum oxide coating created on the sample surface acts as a passive layer and Increases corrosion resistance. However, the effect of the created coating becomes less than the normal case, which can be attributed to the lack of surface smoothness and non-uniformity in the coatings created with electrolytic plasma.

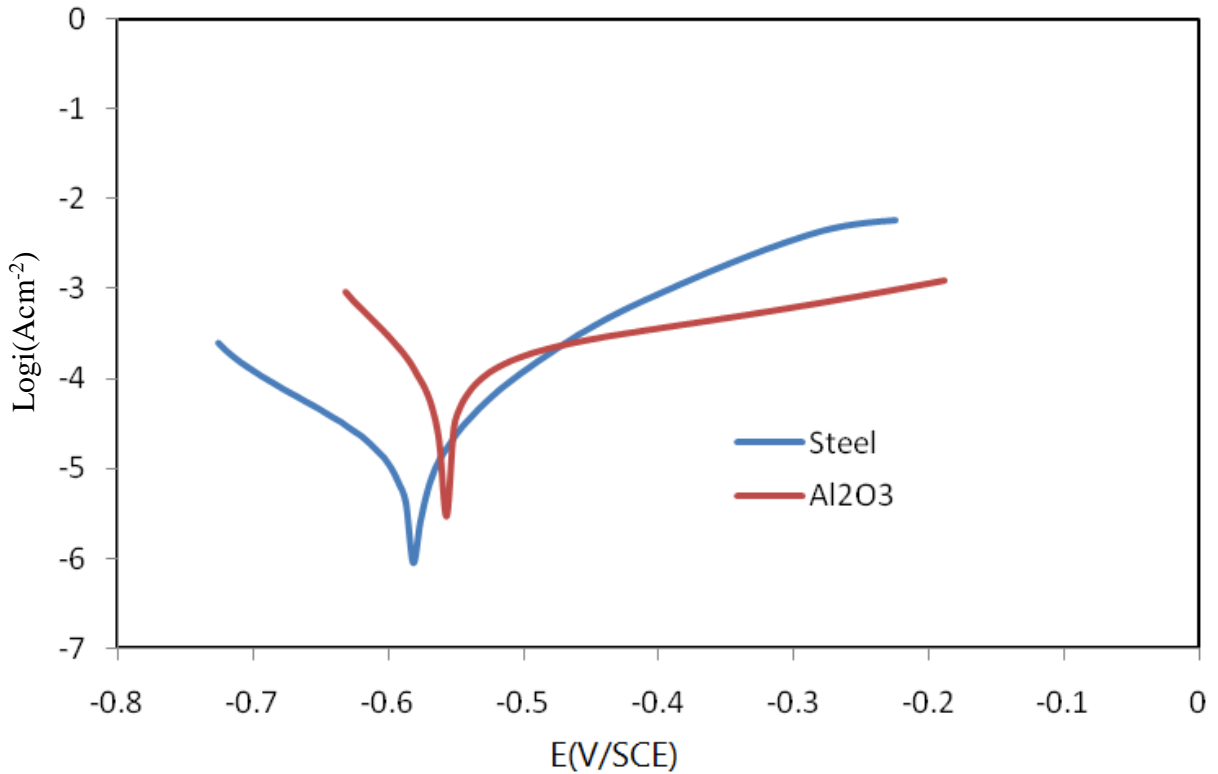

Figure 6. Corrosion diagram of normal and coated sample

**Conclusion**

The created coating is of alpha alumina type, which has a lot of noise due to the lack of uniformity and smoothness of the X-ray diffraction test surface. The created coating has a cauliflower morphology where the size of the grains at the edge of the sample is larger than the middle of the sample. The created coating is like a passive layer and increases corrosion resistance.